\documentclass[aps,prl,twocolumn,floatfix,superscriptaddress,showpacs]{revtex4-1}

\usepackage[dvipdfmx]{graphicx}
\usepackage[dvipdfmx]{animate}
\usepackage{mathrsfs} 
\usepackage[mathscr]{euscript}
\usepackage{amsmath,amssymb}
\usepackage{bm}
\usepackage{color}
\usepackage[vcentermath]{youngtab}
\usepackage{ulem}

\def\ket#1{ | #1 \rangle }
\def\bra#1{ \langle #1 | }

\newcommand{\1}{\mbox{1}\hspace{-0.25em}\mbox{l}} %

\begin{document}


\title{Emergent spin-$1$ Haldane gap and ferroelectricity in a frustrated spin-$1/2$ ladder}

\author{H.~Ueda}
\affiliation{Computational Materials Science Research Team, RIKEN Center for Computational Science (R-CCS), Kobe 650-0047, Japan}
\affiliation{JST, PRESTO, Kawaguchi, Saitama, 332-0012, Japan}

\author{S.~Onoda}
\affiliation{Condensed Matter Theory Laboratory, RIKEN, Wako, Saitama 351-0198, Japan}
\affiliation{Quantum Matter Theory Research Team, RIKEN Center for Emergent Matter Science (CEMS), Wako 351-0198, Japan}

\author{Y.~Yamaguchi}
\affiliation{Division of Materials Physics, Graduate School of Engineering Science, Osaka University, Toyonaka, Osaka 560-8531, Japan}

\author{T. Kimura}
\affiliation{Department of Advanced Materials Science, Graduate School of Frontier Sciences, The University of Tokyo, Kashiwa 277-8561, Japan}

\author{D.~Yoshizawa}
\author{T.~Morioka}
\author{M.~Hagiwara}
\affiliation{Center for Advanced High Magnetic Field Science, Graduate School of Science, Osaka University, Toyonaka, Osaka 560-0043, Japan}
 
 \author{M.~Hagihala}
 \author{M.~Soda}
\author{T.~Masuda}
\author{T.~Sakakibara}
\affiliation{Institute for Solid State Physics, The University of Tokyo, Kashiwa 277-8581, Japan} 

\author{K.~Tomiyasu}
\affiliation{Department of Physics, Tohoku University, Sendai 980-8578, Japan} 
 
\author{S.~Ohira-Kawamura}
\author{K.~Nakajima}
\author{R.~Kajimoto}
\author{M.~Nakamura}
\author{Y.~Inamura} 
\affiliation{Materials and Life Science Division, J-PARC Center, Japan Atomic Energy Agency, Tokai 319-1195, Japan} 

\author{N.~Reynolds}
\affiliation{Laboratory for Neutron Scattering, Paul Scherrer Institut, CH-5232 Villigen, Switzerland}
\author{M.~Frontzek}
\affiliation{Laboratory for Neutron Scattering, Paul Scherrer Institut, CH-5232 Villigen, Switzerland}
\affiliation{Neutron Scattering Division, Oak Ridge National Laboratory, Oak Ridge, TN 37831, USA}
\author{J.~S.~White}
\affiliation{Laboratory for Neutron Scattering, Paul Scherrer Institut, CH-5232 Villigen, Switzerland}

\author{M.~Hase}
\affiliation{National Institute for Materials Science (NIMS), Tsukuba, Ibaraki 305-0047, Japan}

\author{Y.~Yasui}
\affiliation{Department of Physics, School of Science and Technology, Meiji University, Higashi-mita, Tama-ku, Kawasaki 214-8571, Japan}

\begin{abstract}
  We report experimental and theoretical evidence that Rb$_2$Cu$_2$Mo$_3$O$_{12}$ has a nonmagnetic tetramer ground state of a two-leg ladder comprising antiferromagnetically coupled frustrated spin-$1/2$ chains and exhibits a Haldane spin gap of emergent spin-1 pairs. Three spin excitations split from the spin-1 triplet by a Dzyaloshinskii-Moriya interaction are identified in inelastic neutron-scattering and electron spin resonance spectra. A tiny magnetic field generates ferroelectricity without closing the spin gap, indicating a novel class of ferroelectricity induced by a vector spin chirality order. 
\end{abstract}


\maketitle
Quantum spin fluctuations offer a source of various nontrivial states including resonating valence bonds and quantum spin liquids~\cite{anderson:73}. 
In the one-dimensional (1D) antiferromagnet having only the first-neighbor exchange coupling $J_1$ (Fig.~1(a)), the spin quantum number $S$ critically determines the magnitude of quantum spin fluctuations of a long-wavelength mode $\bm{n}(\tau,x)$ around a short-range N\'{e}el order. The topological Berry-phase term gives a contribution of $S^{\mathrm{B}}_{\mathrm{1DHAF}}=i2\pi S Q$ to a nonlinear-$\sigma$ model action for $\bm{n}$ with a topological integer
$Q=\frac{1}{4\pi}\int_0^{1/T}\!d\tau\int\!dx\,\bm{n}\cdot\left(\frac{\partial\bm{n}}{\partial x}\times\frac{\partial\bm{n}}{\partial \tau}\right)$ and the temperature $T$. 
Thus, $e^{-S^{\mathrm{B}}_{\mathrm{1DHAF}}}$ can take $-1$ for a half-integer $S$, allowing for gapless excitations from a disordered ground state. On the other hand, it is always unity for an integer $S$, leading to a so-called Haldane gap~\cite{HALDANE1983464,PhysRevLett.50.1153} in the $S=1$ excitation spectrum from a nonmagnetic ground state~\cite{LIEB1961407,Affleck_review_1989}, as experimentally evidenced in NENP~\cite{0295-5075-3-8-013,Katsumata89,PhysRevLett.65.3181}.
\begin{figure}
\begin{center}
\includegraphics[width=8.7cm]{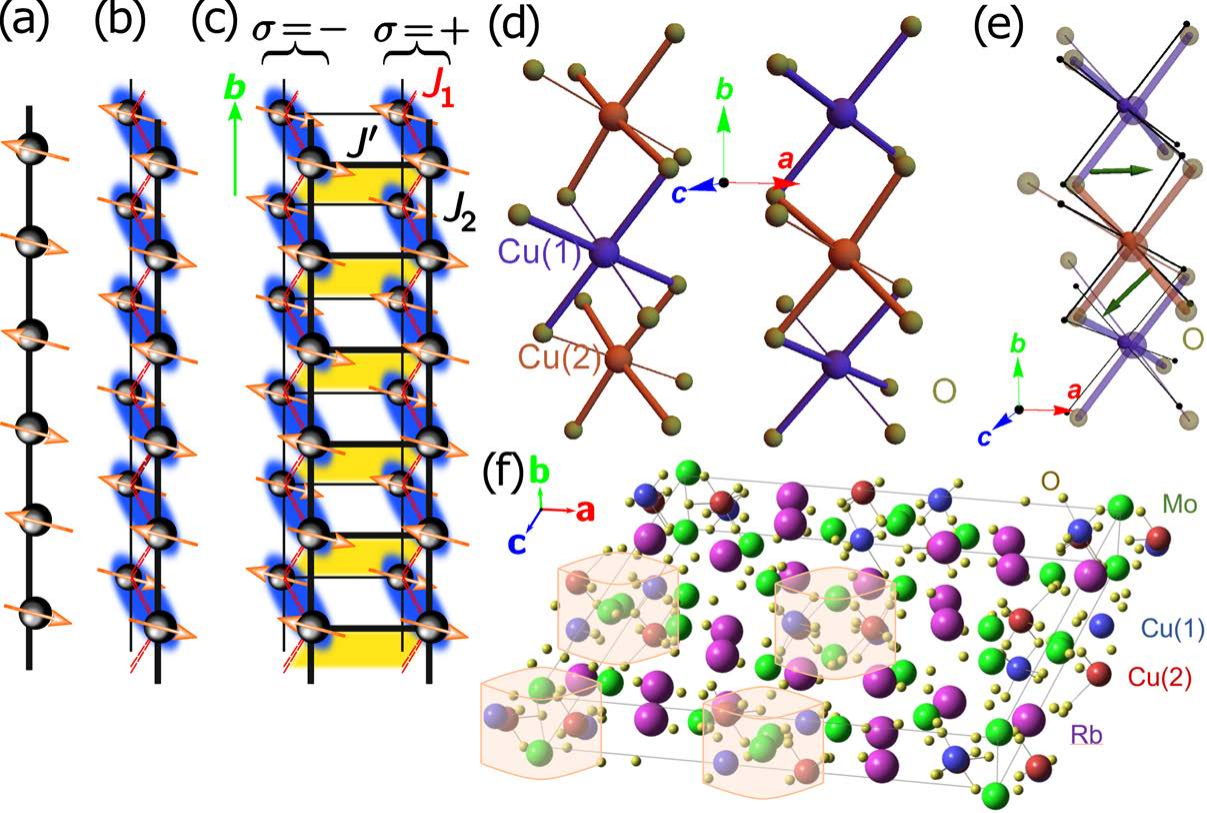}
\end{center}
\caption{
Structures of spin chains.
(a) Antiferromagnetic spin chain.
(b) Frustrated spin-$1/2$ chain with emergent spin-1 pairs (blue clouds). Solid/broken lines represent antiferromagnetic/ferromagnetic interactions. 
(c) Short-range resonating valence bond state involving tetramers (yellow plaquettes) connecting emergent spin-1 pairs. 
(d) Crystal structure of a pair of spin-$1/2$ chains comprising edge-sharing distorted CuO$_6$ octahedra in Rb$_2$Cu$_2$Mo$_3$O$_{12}$.
(e) An ideal centrosymmetric chain of edge-sharing regular CuO$_6$ octahedra (black points), compared with the noncentrosymmetric one in Rb$_2$Cu$_2$Mo$_3$O$_{12}$. Electric dipole moments due to ionic displacements are shown on the first-neighbor Cu spin pairs by green arrows.
(f) A unit cell of Rb$_2$Cu$_2$Mo$_3$O$_{12}$. Two-leg ladders are located in translucent orange tubes.
}
\end{figure}

In the presence of an antiferromagnetic second-neighbor exchange coupling $J_2$, however, the above simple arguments no longer hold. In particular, quasi-1D spin-$1/2$ multiferroic and/or magnetoelectric edge-sharing cuprates, such as LiCu$_2$O$_2$~\cite{Masuda05,Park07}, LiCuVO$_4$~\cite{Enderle05,Naito07}, PbCuSO$_4$(OH)$_2$~\cite{Yasui11,Wolter12}, and Rb$_2$Cu$_2$Mo$_3$O$_{12}$~\cite{Hase04,Yasui13,Yasui14}, involve a ferromagnetic $J_1$ because of nearly 90$^\circ$ Cu-O-Cu bond angles, in addition to an antiferromagnetic second-neighbor exchange coupling $J_2$.
The $J_1$-$J_2$ frustrated spin-$1/2$ Heisenberg chain accommodates a dimerized spin-singlet short-range resonating valence bond ground state~\cite{Itoi01,Furukawa10}. This state is formed by emergent spin-1 pairs (Fig.~1(b)) and has an extremely small Haldane gap and incommensurate short-range spin correlations. Weak but finite easy-plane exchange magnetic anisotropy then induces a quasi-long-range gapless incommensurate spin-spiral and long-range vector spin chirality ($\sum_\ell\langle \bm{S}_\ell\times\bm{S}_{\ell+1}\rangle$)~\cite{Villain} orders~\cite{Nersesyan98,Furukawa10}. A coexisting phase of the vector spin chirality order and the Haldane gap also appears in between the two phases~\cite{Furukawa12}. These states are, however, readily driven to a long-range spiral magnetic order by three-dimensional couplings. This scenario elucidates the ferroelectricity due to the cycloidal magnetism in LiCu$_2$O$_2$~\cite{Park07}, LiCuVO$_4$~\cite{Naito07}, and PbCuSO$_4$(OH)$_2$~\cite{Yasui11}.

In fact, the ferroelectricity associated with the vector spin chirality order may appear robustly in the vector-chiral Haldane dimer phase without the long-range spiral magnetism, if the spin gap is enhanced~\cite{UO14} so that it dominates over the interchain interactions. Indeed, Rb$_2$Cu$_2$Mo$_3$O$_{12}$ provides a unique example of a field-induced ferroelectricity hosted by a nonmagnetic ground state with a spin gap~\cite{Yasui13,Yasui14}. A recent $\mu$SR study also indicates the formation of a spin-singlet state on cooling below $\sim 7$~K and the saturation at around 1-2 K~\cite{Ohira-Kawamura2017}.
In this Letter, we report combined experimental and theoretical evidence that in the quasi-1D cuprate Rb$_2$Cu$_2$Mo$_3$O$_{12}$, a Haldane-gap ground state formed by emergent spin-1 pairs of $S=1/2$ Cu spins (Fig.~1(c)) harbours a ferroelectricity stabilized by a tiny magnetic field. 

Figures~2 shows a temperature dependence of thermodynamic properties of polycrystalline Rb$_{2}$Cu$_{2}$Mo$_{3}$O$_{12}$ samples. The dielectric constant $\varepsilon$ gradually increases on cooling. Then, as in most of magnetically induced ferroelectrics, it exhibits a kink for $B=0.3$ and 0.5~T or a peak for $B=1$ and 2~T at around 8~K (Fig.~2(a)), below which the electric polarization $P$ emerges at an even weaker magnetic field $B=0.05$~T (Fig.~2(b)). Thus, the anomaly in $\varepsilon$ at $B\ge0.05$~T should be ascribed to a ferroelectric transition at $T_{\mathrm{FE}}\sim8$~K. It is natural to expect that the ferroelectric polarization persists at $T<2$~K because of no sign of a reentrant behavior in $\varepsilon$ and $P$ in the low temperature range. Remarkably, $\varepsilon$ does not show a significant decay on cooling down to 2~K for $B\le0.5$~T, while it does for $B\ge1$~T. 
Furthermore, doping nonmagnetic Zn impurities into Cu sites by 2$\%$~\cite{YasuiZn} drastically suppresses $\varepsilon$ and removes the anomaly associated with the ferroelectric transition (Fig.~2(a)).
Therefore, it is clear that the ferroelectricity is indeed triggered by a coherence in the spin degrees of freedom under the weak magnetic field. 
\begin{figure}
\begin{center}
\includegraphics[width=7cm]{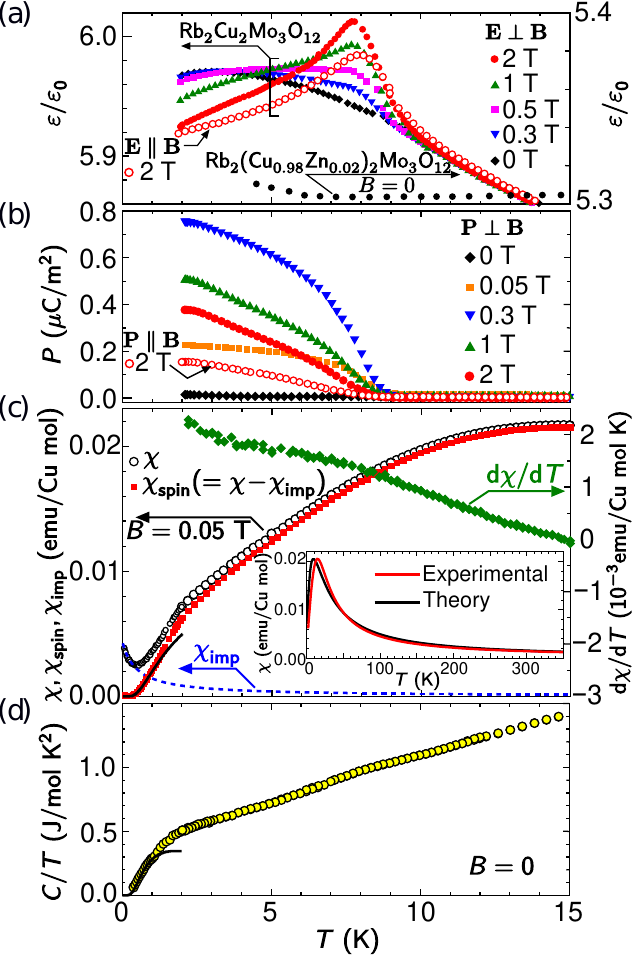}
\end{center}
\caption{
Temperature dependence of thermodynamic properties of polycrystalline Rb$_2$Cu$_2$Mo$_3$O$_{12}$.
(a) Dielectric constant $\varepsilon/\varepsilon_0$.
(b) Electric polarization $P$ at magnetic fields. Note that a powder average of the magnetic field direction broadens the transition.
(c) Magnetic susceptibility $\chi$ (open circles). The impurity contribution $\chi_{\mathrm{imp}}$, responsible for the upturn of $\chi$ below 0.5~K, was fitted by the Curie-Weiss law with the spin vacancy  concentration of $0.5~\%$ and the Weiss temperature $-0.5$~K (blue dashed curve). Red points represent the data $\chi_{\mathrm{spin}}$ subtracted by $\chi_{\mathrm{imp}}$.  Also shown is $d\chi/dT$ (green points). The inset shows a high-temperature fitting of $\chi$ (black curve) with a powder average of the exact diagonalization results (red curve).
(d) Specific heat $C$ at $B=0$. The solid curves in (c) and (d) are the fitting curves proportional to $\exp(-E_g/T)$ with the energy gap $E_g = 1.7$~K. 
}
\end{figure}
\begin{figure}
\begin{center}
\includegraphics[width=6cm]{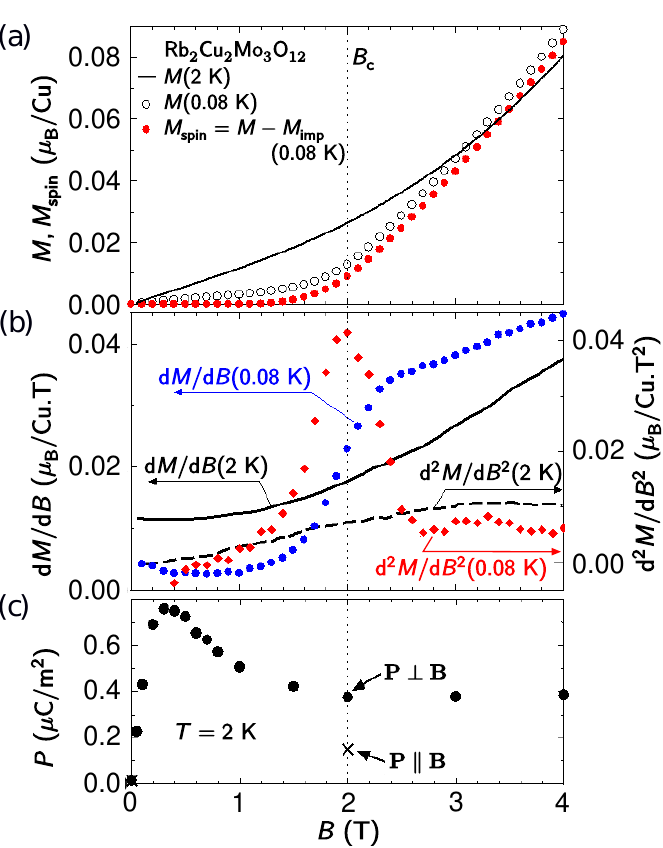}
\end{center}
\caption{
Magnetic field dependence of thermodynamic properties of polycrystalline Rb$_2$Cu$_2$Mo$_3$O$_{12}$.
(a) Magnetization $M$ per Cu atom. 
(b) Derivatives of $M$ with respect to $B$. 
(c) Ferroelectric polarization $P$ for both $\bm{P}\parallel\bm{B}$ and $\bm{P}\perp\bm{B}$. 
}
\end{figure}

The signals of both $\varepsilon$ and $P$ below $T_{\mathrm{FE}}$ are larger for the configuration of $\bm{E},\bm{P}\perp\bm{B}$ than for $\bm{E},\bm{P}\parallel\bm{B}$ at least at 2~T (Figs.~2(a) and 2(b)), as in many edge-sharing multiferroic cuprates showing a cycloidal magnetic order~\cite{Park07,Naito07,Yasui11}. This implies that the uniform vector spin chirality gives rise to a dominant contribution to the ferroelectric polarization among many mechanisms~\cite{Jia07}. On the other hand, no anomaly is observed in the magnetic susceptibility $\chi$ and $d\chi/dT$ (Fig.~2(c)), in contrast to the multiferroic cuprates~\cite{Park07,Naito07,Yasui11}. Moreover, a spin gap $E_g\sim1.7$~K has been observed in both $\chi$ and the specific heat $C$~\cite{Yasui13,Yasui14} (Figs.~2(c) and 2(d)).

The emergence of this spin gap is also confirmed by the measurements of the magnetization $M$.
Figures~3(a) and 3(b) present experimental results on $M$, and $dM/dB$ and $d^2M/dB^2$, respectively. A subtraction of a small impurity contribution as outlined in Fig.~2 caption reveals that $M$ at $T=0.08$~K shows an activation by the threshold field $B_c\sim2.0$~T where $d^2M/dB^2$ exhibits a peak.
On the other hand, at a much lower temperature $T=2.0$~K than $T_{\mathrm{FE}}$, $\bm{P}(\perp \bm{B})$ steeply appears at a much lower field, at least 0.03~T, than $B_c$ (Fig.~3(c)). It exhibits a broad peak at around 0.2-0.3~T, and then gradually decays to a constant at higher fields up to 4~T. This observation confirms that the ferroelectricity is stabilized by a tiny magnetic field but not affected by a closing of the spin gap and an onset of the magnetization at $B_c$. Namely, at the energy scale associated with 0.03~T or less, there exists
a low-energy mode, which is magnetic-dipole inactive but electric-dipole
active, and thus linearly coupled to the vector spin chirality.

\begin{figure}
\begin{center}
\includegraphics[width=8.7cm]{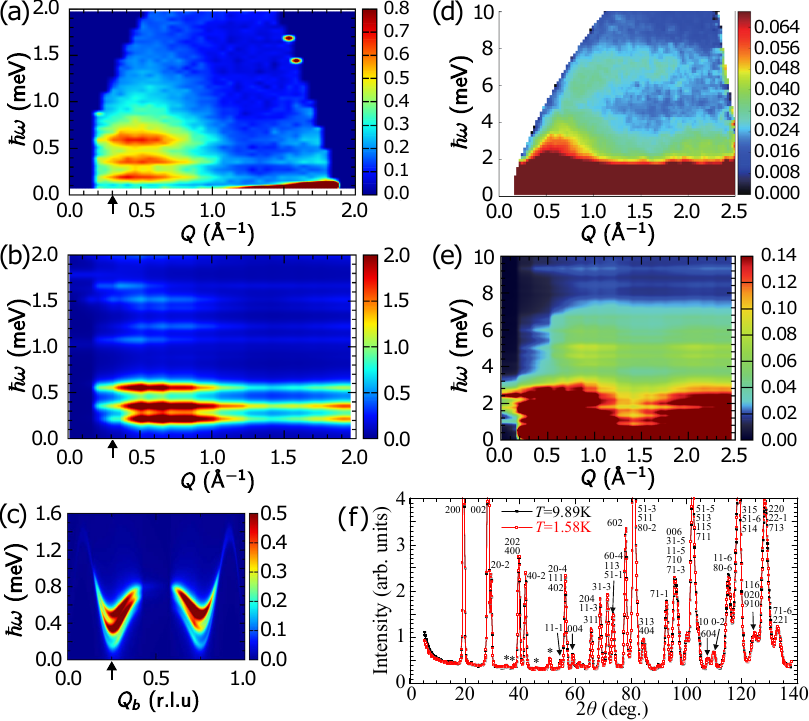}
\end{center}
\caption{
    Neutron-scattering results on polycrystalline
     Rb$_2$Cu$_2$Mo$_3$O$_{12}$ at $B=0$.
(a) Experimental and (b) theoretical low-energy powder-averaged spectra, measured at 1.6~K and calculated at $T=0$, respectively.
(c) Theoretical low-energy spectra along the $b$ axis without the powder average. The results obtained from a 28-site cluster by taking the parameter set for the thermodynamic limit~\cite{supplement} have been interpolated.
The incommensurate wavevector $Q=0.3$~\AA$^{-1}$ is marked by black arrows in (a), (b) and (c).
(d) Experimental and (e) theoretical powder-averaged spectra, measured at 6.5~K and calculated at $T=0$, respectively, in a wider energy range. Note that the incommensurate wavevector is shifted downwards from the maximum position of the powder-averaged spectra to the onset in the panels (a) and (b). (f) Neutron powder diffraction patterns of Rb$_2$Cu$_2$Mo$_3$O$_{12}$ measured at 9.89~K$>T_{\mathrm{FE}}$ (black) and 1.58~K$<T_{\mathrm{FE}}$ (red) in $B=0$~T. The four peaks with $*$ symbols are derived from a nonmagnetic impurity phase Rb$_2$Mo$_3$O$_{10}$. A cold neutron wavelength $\lambda = 4.5\AA$ was chosen.
}
\end{figure}
All the above thermodynamic properties provide evidence of a spin-gapped ferroelectric behavior stabilized by the tiny applied magnetic field, most likely through the vector spin chirality. It should also be possible to confirm this from spectral properties. To probe $S=1$ triplet excitations from the nonmagnetic ground state, low-energy inelastic neutron-scattering experiments have been performed on powder samples.
Figure~4(a) represents the results at 1.5~K measured on the AMATERAS spectrometer. Discrete excited levels are clearly seen at 0.2, 0.38, and 0.6~meV. 
The periodicity of these spin excitations along the chain can be determined from the onset wavenumber $Q \sim 0.3$ \AA$^{-1}$ of the powder-averaged intensities, and roughly corresponds to eight spins.
A natural interpretation will be that  $S=1$ triplet excitations are split into the three by Dzyaloshinskii-Moriya interactions.
Note that cooling below $T_{\mathrm{FE}}$ and applying the magnetic field do not alter the diffraction patterns (Fig.~4(f) and Ref.~\cite{Reynolds_2019}): neither a superlattice peak nor any visible additional diffraction intensity appears.
Note also that a clear long-range magnetic order is absent at the incommensurate wavevector $(0, Q_b, 0)$ in the accuracy of 0.06 $\mu_B$.　
Actually, the absence of clear muon spin precession/relaxation~\cite{Ohira-Kawamura2017} precludes a long-range order of all the Cu spins with the moment amplitude $\gtrsim0.01\ \mu_B$ and of dilute ($>1\%$) Cu or impurity spins with the moment amplitude 1~$\mu_B$. The possibility of having a tiny fraction ($<1\%$) of magnetically ordered domains in the polycrystalline samples might be hardly ruled out. However, such order is absolutely extrinsic and irrelevant to the observed magnetic and ferroelectric properties of the bulk, because the Weiss temperature $-0.5$~K is much lower than $T_{\mathrm{FE}}$ and the exchange coupling constants obtained below.

The overall experimental results on the magnetic properties can be elucidated theoretically from the following two-leg ladder model of frustrated $J_1$-$J_2$ spin-1/2 chains (Fig.1(c))~\cite{supplement}
\begin{eqnarray}
  H&=&
  \sum_\ell\sum_{\sigma=\pm}\Big[
    \sum_{j=1,2} J_j \bm{S}_{\sigma,\ell}\cdot\bm{S}_{\sigma,\ell+j}
    +J'\bm{S}_{+,\ell}\cdot\bm{S}_{-,\ell}, \nonumber\\
    && +\sigma\left((-1)^\ell \bm{D}_{\rm s} \cdot \bm{S}_{\sigma,\ell} \times \bm{S}_{\sigma,\ell+1}
    + \bm{D}_{\rm u} \cdot \bm{S}_{\sigma,\ell}\times\bm{S}_{\sigma,\ell+1}\right) \nonumber\\
    && -g\mu_B\bm{B}\cdot\bm{S}_{\sigma,\ell} \Big]
  \label{eq:H}
\end{eqnarray}
with the $g$-factor $g=2.16$~\cite{supplement} and the applied magnetic field $\bm{B}$, where $\bm{S}_{\sigma,\ell}$ stands for an $S=1/2$ spin at the site $\ell$ in the chain of edge-shared CuO$_6$ octahedra (Fig.~1(d)) labeled by the index $\sigma=\pm$.
It has already been revealed that the antiferromagnetic rung exchange coupling $J'$ between the nearest-neighbor spins in the adjacent $J_1$-$J_2$ chains is required for enhancing the spin gap~\cite{HUSO_2020}.
$\bm{D}_{\mathrm{u}}$ and $\bm{D}_{\mathrm{s}}$ represent the uniform and staggered components of intrachain Dzyaloshinskii-Moriya vectors caused by two inequivalent first-neighbor Cu-Cu bonds involving noncollinear electric dipole moments as shown by green arrows in Fig.~1(e). No crystal symmetry constrains the directions of the Dzyaloshinskii-Moriya vectors. However, since the numerical results shown below are insensitive to a nonzero value of $\bm{D}_{\mathrm{u}}\cdot\bm{D}_{\mathrm{s}}$, we take $\bm{D}_{\mathrm{u}}\perp\bm{D}_{\mathrm{s}}$.
Henceforth, we adopt $J_1=-114$~K,  $J_2=35.1$~K, $J'=20.5$~K, $D_{\rm s}=44.3$~K, and $D_{\rm u}=24.4$~K to explain overall results of the magnetic susceptibility and inelastic neutron-scattering spectra
from exact-diagonalization calculations on a 16-site cluster. (See Supplemental Material~\cite{supplement} for examinations of finite-size effects by means of the density-matrix renormalization group for infinite systems (iDMRG).)
Indeed, the numerical results on $\chi$ for $B=0$ reasonably agree with the experimental data~\cite{Hase04}, as shown in the inset of Fig.~2(c), and the iDMRG result $2.15$ T on the critical magnetic field agrees with the experimental one $\sim2.0$ T (Fig.~3(b)). 
Furthermore, the experimental results on the low-energy powder-averaged inelastic neutron-scattering spectra (Fig.~4(a)) are nicely explained by the theoretical results (Fig.~4(b))~\cite{supplement}.
From Figs.~4(a) and 4(b), the three low-energy excitations might look dispersionless. However, this is an artifact of powder averaging. Figure~4(c) shows the theoretical results of the dispersive spectra as a function of the particular wavevector component $Q_b$ in the chain direction, the crystallographic $b$-axis, with $Q_a=Q_c=0$.
Actually, the agreement in the inelastic neutron-scattering spectra extends to a much higher energy $\sim10$~meV, as is apparent by comparing the current experimental results in the high energy range measured at 4SEASONS spectrometer (Fig.~4(d)), which are refined from the previous data~\cite{Tomiyasu09}, with the theoretical results~\cite{supplement} (Fig.~4(e)).

The scenario of a splitting of the $S=1$ excited states due to Dzyaloshinskii-Moriya interactions is also supported by electron spin resonance (ESR) experiments on powder samples. 
Figure~5(a) presents the temperature dependence of the ESR transmission spectra at a frequency $f=81~\mathrm{GHz}\sim0.33~\mathrm{meV}$ as a function of $B$. A paramagnetic resonance is found as a significantly broad peak at 2.7~T for a much lower temperature 8.7~K than $J_1$, $J_2$ and $J'$, as indicated by red arrows. It should appear as a much sharper peak in the absence of moderately large Dzyaloshinskii-Moriya interactions~\cite{Zorko04}. On cooling, the peak becomes even more broadened, and eventually bifurcates below 5~K.
In the frequency dependence of the ESR spectra at 1.6~K (Fig.~5(b)), this new low-energy mode (green arrows) has been identified, as well as another lower-energy mode (blue arrows).
These two modes are plotted with $\triangle$ and $\triangledown$ in the $B$-$f$ diagram of Fig.~5(c), in favorable comparison with a density plot of the theoretical results~\cite{supplement} on the optical absorption power~\cite{Slichter90} at the same temperature.
  The dominant contributions to the two series originate from thermally activated transitions. Theoretically, the second-lowest-energy mode ($\triangledown$) is ascribed to transitions from the first excited state to the third at the wavevector $Q_b=1/4$~r.l.u. (the right panel of Fig.~5(d)) and from the first excited state to the second at $Q_b=0$ (the left panel of Fig.~5(d)), as shown by solid and dashed curves in Fig.~5(c), respectively.
  The lowest-energy mode ($\triangle$) is ascribed to transitions from the first excited state to the second and from the second to the third at $Q_b=1/4$~r.l.u. (the right panel of Fig.~5(d)), as shown by two solid curves in Fig.~5(c).
  A significantly large dependence of the excitation energies on the field direction in the theoretical calculations shown in Fig.~5(d) also elucidates the unusually broad spectral features identified in the powder ESR experiments.

\begin{figure}
\begin{center}
\includegraphics[width=8.7cm]{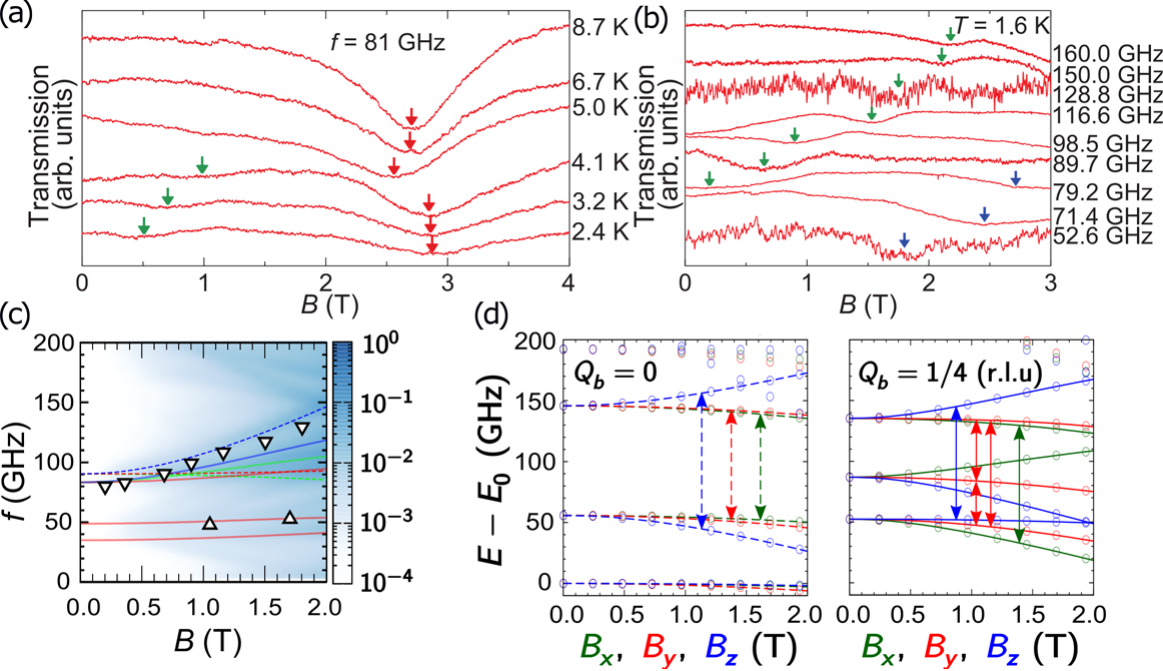}
\end{center}
\caption{
Electron spin resonance spectra of polycrystalline Rb$_2$Cu$_2$Mo$_3$O$_{12}$.
(a) Temperature dependence of the experimental transmission spectra at 81~GHz. Arrows represent the resonance fields.
(b) Experimental transmission spectra at 1.6~K for designated frequencies. Green and blue arrows denote two sequences of resonance fields. 
(c) Theoretical optical absorption power at 1.6~K. Experimentally observed resonance fields indicated by blue and green arrows in (a) are plotted by $\triangle$ and $\triangledown$, respectively, for comparison.
(d) Energy levels at $Q_b=0$ (left) and $Q_b=1/4$~r.l.u. (right) computed under $\bm{B}$ applied along the $x$, $y$, and $z$ directions, where $\bm{D}_u\parallel z$ and $\bm{D}_s\parallel x$. Transitions denoted by the arrows in the left/right panel produce resonance spectra shown by dashed/solid curves in (b).
}
\end{figure}

The current frustrated spin-$1/2$ ladder model, that has reproduced overall experimental results on Rb$_2$Cu$_2$Mo$_3$O$_{12}$, actually has a tetramer-singlet ground state formed by emergent $S=1$ spins with a Haldane gap. (See Fig.~1(c).) This ground state is adiabatically connected to the limit of the two decoupled chains with $J'=0$, each of which has a singlet Haldane dimer ground state~\cite{Furukawa12}, and then to the two decoupled spin-1 Haldane chains, as in an antiferromagnetic spin-1 ladder~\cite{Todo01}.
At present, it remains open to explain the ferroelectricity stabilized by a tiny magnetic field. Nevertheless, it is clear from the symmetry that it is accompanied by a genuine long-range vector spin chirality order, which is not parasitic to a (quasi-)long-range spiral magnetic order. This ground state has long been sought since the proposal by Villain~\cite{Villain}. Thus, the current study uncovers a novel class of magnetically induced ferroelectricity in the absence of a long-range magnetic order, in contrast to many multiferroic magnets due to a cycloidal magnetism.
A quest to more microscopic properties of this novel ferroelectric (vector-spin-chirality ordered) emergent Haldane-gap state will demand experiments on the single crystals and the associated microscopic theoretical analyses in the future.

\begin{acknowledgments}
The authors acknowledge I. Terasaki for his support of the work and helpful discussion and K. Kaneko for preliminary neutron scattering experiments on the triple-axis spectrometer LTAS installed at JRR-3 reactor, Japan. Numerical calculations were partially performed by using the RIKEN Integrated Cluster of Clusters and the RIKEN HOKUSAI supercomputers. The time-of-flight neutron-scattering experiments were performed using the chopper spectrometers AMATERAS and 4SEASONS at J-PARC (Proposal Nos. 2012P0202 and 2009A0093). The work was partially supported by Grants-in-Aid for Scientific Research (Grant 24244059, 24740253, 25220803, 25246006, 25800221, 16K05426, 17001001, 17H06137 and 17K14359) from Japan Society for the promotion of Science and by the RIKEN iTHES project and by the FRIS Program for the creation of interdisciplinary research at Tohoku University and by Swiss National Science Foundation via the SNSF project grant No. 200021\_153451. The neutron-diffraction experiments were performed at the Swiss spallation neutron source SINQ, Paul Scherrer Institute, Villigen, Switzerland. 
H.U. is grateful to S. Yunoki for his support.
\end{acknowledgments}

\bibliographystyle{apsrev4-1}
%

\clearpage
\onecolumngrid
\renewcommand{\thefigure}{S\arabic{figure}}
\renewcommand{\theequation}{S\arabic{equation}}
\setcounter{equation}{0}
\setcounter{figure}{0}

\section{Supplemental Material for ``Emergent spin-$1$ Haldane ladder from the one-dimensional frustrated spin-$1/2$ magnet Rb$_2$Cu$_2$Mo$_3$O$_{12}$''}

\section{Crystal structure of Rb$_2$Cu$_2$Mo$_3$O$_{12}$}
The quasi-one-dimensional spin-$1/2$ frustrated chain compound Rb$_2$Cu$_2$Mo$_3$O$_{12}$ belongs to the space group $C2/c$ above $T_{\mathrm{FE}}\sim8$~K (ref.~\cite{Solodovnikov97}). The unit cell comprises four symmetry-related two-leg spin ladders (Fig.~1(g)). Two spin chains in each ladder are related by an inversion operation. Each spin chain contains two Cu$^{2+}$ sites in the unit cell with a slight alternation of the first-neighbor Cu-Cu distances along the chain direction, namely, the crystallographic $b$ axis~\cite{Solodovnikov97} (Fig.~S1(a)). The first-neighbor Cu-O-Cu bond angles approximate to 90$^\circ$, leading to ferromagnetic first-neighbor exchange couplings. 
A large buckling of CuO$_2$ chains yields the Cu-Cu-Cu bond angle $\theta=111.75^\circ$ along each chain and the direction from the Cu$^{2+}$ ion to the apical oxygen alternates in each chain. These distortions should produce a moderately large magnetic anisotropy in the exchange coupling. The structures shown in Figs.~1(e) and 1(d) of the main text can be viewed from various directions by a rotation about the $b$ axis in Figs.~S1(b) and S1(c).

\begin{figure}
\vspace{-4mm}
{\large \hspace{-25mm} (a) \hspace{50mm} (b) \hspace{50mm} (c)} \par
\vspace{0mm}
\begin{center}
\includegraphics[height=2.0in]{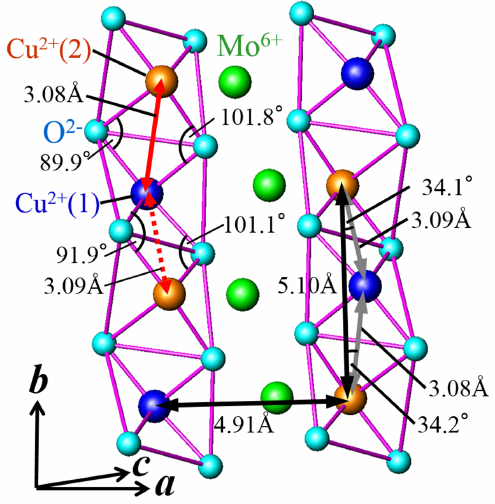}
~~
\animategraphics[height=2.0in,autoplay,controls,buttonsize=1.00em,loop]{12}{Rb2Cu2Mo3O12idealcrystalwoOlabels_}{000}{089}
~~
\animategraphics[height=2.0in,autoplay,controls,buttonsize=1.00em,loop]{12}{Rb2Cu2Mo3O12bothcrystalwoOlabels_}{000}{089}
\end{center}
\caption{ Crystal structure of the spin-$1/2$ ladder in Rb$_2$Cu$_2$Mo$_3$O$_{12}$. we take the following representations of three primitive lattice vectors; $\bm{a}=(26.3844,0,-8.1956)~\mathrm{\AA}$, $\bm{b}=(0,5.1018,0)~\mathrm{\AA}$, and $\bm{c}=(0,0,19.2920)~\mathrm{\AA}$. 
  (a) Crystal structure of Rb$_{2}$Cu$_{2}$Mo$_{3}$O$_{12}$.
  (b) Gif animation of the structure displayed in Fig. 1(e) of the main text.
  (c) Gif animation of the structure displayed in Fig. 1(f) of the main text.
  In (b) and (c), thin lines represent the bonds connected from Cu to the apical oxygens.
}
\label{fig:lattice}
\end{figure}

\section{Determination of the exchange coupling constants}

\subsection{Model.}
Here we construct a plausible model for Rb$_2$Cu$_2$Mo$_3$O$_{12}$. The material contains two inequivalent edge-shared CuO$_6$ octahedra that are aligned alternately in a zigzag manner along the crystallographic $b$ axis, as shown in Fig.~S1. 
Each spin chain couples to its partner transformed by a spatial inversion and separated from each other by 4.91~$\mathrm{\AA}$ (Fig.~1(e)), which is comparable to the second-neighbor intrachain Cu-Cu distance. This suggests a nonnegligible antiferromagnetic interchain coupling $J'$ of the order of $J_2$. Both of the two inequivalent CuO$_6$ octahedra are largely distorted from regular octahedra (Fig.~1(f)). In effect, the directions of apical oxygens from Cu sites and thus the $d$-orbital shapes of unpaired electrons alternate, in sharp contrast to a simple edge-sharing CuO$_6$ octahedral network in many other compounds. This should yield a relatively large Dzyaloshinskii-Moriya interaction, as it is expected to elucidate the experimentally observed splitting of the triplet excitations. 
Thus, as a minimal effective Hamiltonian for describing low-energy magnetic properties of Rb$_2$Cu$_2$Mo$_3$O$_{12}$, we propose a frustrated $J_1$-$J_2$ spin-$1/2$ two-leg ladder model with noncollinear uniform and staggered Dzyaloshinskii-Moriya (DM) interactions, namely,
\begin{eqnarray}
  H&=&H_{\mathrm{SU(2)}}+H_{\mathrm{DM}}+H_{\mathrm{Z}}.
  \label{eq:H}
\end{eqnarray}
Here,
\begin{eqnarray}
  H_{\mathrm{SU(2)}}&=&\sum_\ell \left[\sum_{\sigma=\pm}\sum_{j=1,2} J_j \bm{S}_{\sigma,\ell}\cdot\bm{S}_{\sigma,\ell+j}
  +J'\bm{S}_{+,\ell}\cdot\bm{S}_{-,\ell}\right],
  \label{eq:H_SU(2)}
\end{eqnarray}
is the SU(2)-symmetric exchange interaction part with the ferromagnetic first-neighbor coupling $J_1<0$, the antiferromagnetic second-neighbor coupling $J_2>0$, and the antiferromagnetic interchain rung coupling $J'>0$, while
\begin{eqnarray}
  H_{\mathrm{DM}}&=&\sum_\ell \sum_{\sigma=\pm}\sigma 
  \left[(-1)^\ell \bm{D}_{\mathrm{s}} \cdot \left(\bm{S}_{\sigma,\ell} \times \bm{S}_{\sigma,\ell+1}\right)
    + \bm{D}_{\mathrm{u}} \cdot \left(\bm{S}_{\sigma,\ell}\times\bm{S}_{\sigma,\ell+1}\right) \right],
  \label{eq:H_DM}
\end{eqnarray}
is the DM interaction part with the DM vectors $\bm{D}_{\mathrm{u}}$ and $\bm{D}_{\mathrm{s}}$ for the uniform and staggered components, respectively. 
A magnetic field $\bm{B}$ is introduced through the Zeeman term
\begin{eqnarray}
  H_{\mathrm{Z}}&=&-g\mu_B\bm{B}\cdot\sum_{\sigma=\pm}\sum_\ell \bm{S}_{\sigma,\ell},
\end{eqnarray}
with the $g$ factor $g=2.16$ (Fig.~\ref{fig:esr_g}). The relation between the spin coordinate and crystallographic coordinate frames will be fixed later.

Now, before explaining how the model parameters are determined, we give some comments on an issue of a single chain versus a two-leg ladder. In the previous papers~\cite{UO13,UO14,Agrapidis_2017}, a single spin-$1/2$ $J_1$-$J_2$ XXZ chain has been studied by taking into account a small/large bond alternation in the amplitude of $J_1$, which is crystallographically present in the material. Then, it has been found that an infinitesimally small bond alternation drastically changes the phase diagram: a vector-spin-chirality ordered phase showing quasi-long-range gapless incommensurate spin correlations in the case with a moderately large easy-plane exchange anisotropy~\cite{Furukawa10} is split into two vector-spin-chirality ordered phases showing only short-range gapped spin correlations~\cite{UO13}. Actually, the two phases are topologically distinct in the presence of the time-reversal symmetry and separated by a single critical line~\cite{UO14}. A possibility of explaining the experimental results for Rb$_2$Cu$_2$Mo$_3$O$_{12}$ in terms of an either phase has then been examined, before proceeding to detailed analyses of the two-leg ladder model given by Eq.~(\ref{eq:H}). However, we have found that neither is the case~\cite{HUSO_2020}, as we will explain below.

The dimer phases with less easy-plane XXZ anisotropy has a too small spin gap, whether they are accompanied by a vector-spin-chirality order or not. If we enhance the spin gap by increasing the bond alternation to an unphysically large value, as discussed in Ref.~\cite{Agrapidis_2017}, or by tuning the antiferromagnetic coupling $J_2$, as discussed in Ref.~\cite{Agrapidis_2018}, then the incommensurate wavevector of spin correlations approaches a periodicity of four spins, in contrast to eight spins in experiments. 
The alternating strength of $J_1$ is too small since the first-neighbor bond length alternates only by 0.3\% and 0.3\% of $J_1$ is an order of magnitude smaller than the experimentally observed energy gap of 0.2 meV. The only practical effect of the alternation in $J_1$ is that it eliminates the twofold topological degeneracy associated with two choices of forming a plaquette singlet RVB. But this topological degeneracy does not manifest itself in the bulk properties measured in the current experiments at all.
The other dimer phases, with and without the vector-spin-chirality order, in the case of strong easy-plane anisotropy may have a large spin gap compared to the experiment, but then the initial slope of the magnetization curve, namely, the magnetic susceptibility, within the easy-plane directions becomes an order of magnitude larger than in the experimental results. Furthermore, replacing the XXZ exchange anisotropy with DM interactions does not improve the case. Thus, we conclude that a different interaction should dominantly enhance the spin gap.

Taking into account the alternation in $J_2$ and any more symmetry-allowed spin-spin interactions does not change the symmetry of our model Eq.~(\ref{eq:H}). In fact, from quantitative viewpoints, the amplitude of the symmetry-allowed alternation in $J_2$ should be extremely small compared to $J_2$ itself, since the two bond lengths between Cu(1) sites and between Cu(2) sites are equivalent and the associated exchange paths are also similar. Also from qualitative viewpoints, the alternation in $J_2$ does not produce any new frustration effects either. Therefore, it is a very good starting point to neglect the alternation in $J_2$. Effects of the DM interactions between the second-neighbor spins, including their alternation, are clearly even more subdominant. The reason why the DM interaction between the first-neighbor spins is rather large compared to other cuprates is that the directions of apical oxygens of CuO$_6$ octahedra alternate largely. This is not the case for the second-neighbor Cu spins. The possible rung DM interaction should not also be large since it is mediated by empty 5s and 4d electrons of Mo ions. It does not produce global effects either, since they alternate in the chain direction by symmetry and so the effects are cancelling each other.

Therefore, the most natural candidate is the antiferromagnetic interchain rung interaction $J'$, as we have already explained. In fact, when $J'$ and $D_{\mathrm{s}}$ are both finite, the small bond alternation in $J_1$ is found to have negligibly small effects and thus has been left out.

\subsection{SU(2)-symmetric exchange coupling constants.}

First, we consider the predominant SU(2)-symmetric part $H_{\mathrm{SU(2)}}$, to adjust $J_1/J_2$ and $J'/J_2$. We numerically compute the spin gap between the ground state in the $S=0$ sector and the sixfold degenerate first excited states in the $S=1$ sector as well as equal-time spin-spin correlation functions in the ground state by means of the infinite-size density matrix renormalization group (iDMRG) technique~\cite{white92,white93,McCulloch08}.
In particular, a four-spin unit is adopted for the matrix-product state whose matrix dimension is taken up to 800 to ensure the convergence.

The spin gap and the maximum peak position $q_0$ in the Fourier transform 
\begin{equation}
  S^{x,x}(q)= \sum_{\ell=1}^{N} \langle S^{x}_{+,0} S^{x}_{+,\ell} \rangle
  \exp\left( - i \frac{q\ell b}{2N}\right)
\label{eq:fourier_component}
\end{equation}
with $N=256$ and half the lattice constant $b/2=2.55$~\AA~are plotted as functions of $J_1/J_2$ and $J'/J_2$ in Figs.~S2(a) and S2(b), respectively. In the experimental data of the inelastic neutron scattering spectrum shown in Fig.~4(a) of the main text, the intensities of three dispersionless spin excitation levels that are located at $\omega=0.20$, $0.38$, and $0.60$~meV rise at around $Q_b \sim 0.3$~\AA$^{-1}$. This value of the incommensurate wavevector gives a constraint that $J_1/J_2$ and $J'/J_2$ should lie around a boundary between blue area and green region in the numerical results shown in Fig.~S2(b). Note that the first excited $S=1$ states in the SU(2)-symmetric case are split and largely affected by the DM interaction, as we will show below. On the contrary, the high-temperature magnetic susceptibility is insensitive to the DM interactions. Indeed, under the above constraint $q_0\sim0.3$~\AA$^{-1}\sim0.25$~r.l.u., we have succeeded in fitting the experimental results in the inset of Fig.~2(c) reasonably well by the numerical exact-diagonalization results on a 16-site cluster for a choice of $J_1/J_2=-3.24$ and $J'/J_2=0.583$, and $J_2=34.4$~K if we take into account the sum $\chi_0$ of the van Vleck and diamagnetic susceptibilities (Fig.~S3(a)), as shown in Fig.~S3(b).

\begin{figure}
\begin{center}
\includegraphics[width=8.6cm]{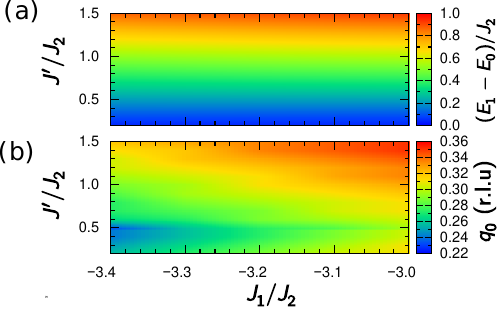}
\end{center}
\caption{ Spin gap and incommensurability in the SU(2)-symmetric model.
  (a) Spin gap between the ground state with $S=0$ and the first excited states with $S=1$.
  (b) Maximum peak position $q_0$ in the Fourier component $S^{x}(q)$ of Eq.~(\ref{eq:fourier_component}) as a function of $J_1/J_2$ and $J'/J_2$ for the SU(2)-symmetric Hamiltonian $H_{\mathrm{SU(2)}}$.
}
\label{fig:gap_pitch}
\end{figure}

\begin{figure}
\begin{center}
\includegraphics[width=8.6cm]{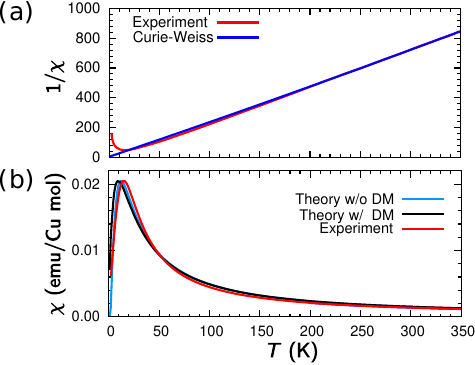}
\end{center}
\caption{
(a) Curie-Weiss fitting of $\chi=(g\mu_{\rm B}/2)^2/(T-\Theta)+\chi_0$ for the magnetic susceptibility $\chi$ of $B=0$ in the scale of $1/\chi$ with $g=2.16$ and the Bohr magneton $\mu_{\rm B}$, where the best fit has been obtained in a temperature range $T=[200,350]$ K for $\Theta=-2.3$ K and the sum $\chi_0=-6.1\times10^{-5}$ (emu/Cu mol) of the van Vleck and diamagnetic susceptibilities. (b) Temperature dependence of $\chi$. It has been calculated by the numerical diagonalization of the 16-site cluster. The blue and black curves represent the numerical results for $H_{\rm SU(2)}$  with  $J_2=34.4$ K and for $H_{\rm SU(2)} +H_{\rm DM}$ with $J_2=35.1$ K, $D_u/|J_1|=0.215$, and $D_s/|J_1|=0.39$. In the both cases, $J_1/J_2=-3.24$ and $J'/J_2=0.583$ are commonly taken.
}
\label{fig:chi}
\end{figure}

\subsection{DM coupling constants.}
Next, we fix the ratios of $J_1/J_2=-3.24$ and $J'/J_2=0.583$, as have been determined above. Then, we adjust the DM coupling constants as well as $J_2$ so that the first excited $S=1$ levels at $Q_b \sim q_0$ in the SU(2)-symmetric case are split into three located at 0.2, 0.38, and 0.6~meV, as in experimental observations. Note that the magnetic susceptibility and the incommensurability examined above are almost intact. 
We have found that two types of DM interactions dominantly control the level splitting: a uniform DM interaction $D_{\rm u}$ and a staggered DM interaction $D_s$ with their DM vectors $\bm{D}_{\mathrm{u}}$ and $\bm{D}_{\mathrm{s}}$ being perpendicular to each other. To be explicit, we henceforth take $\bm{D}_{\mathrm s} \parallel x$ and $\bm{D}_{\mathrm u} \parallel z$.

\begin{figure}
\begin{center}
\includegraphics[width=16.0cm]{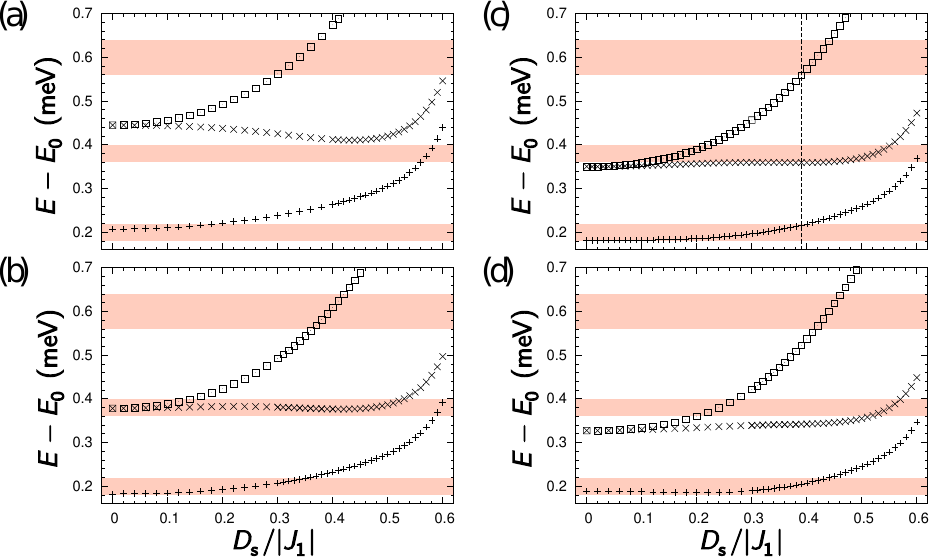}
\end{center}
\caption{Three lowest-energy levels with $q_0=1/4$ r.l.u. calculated by numerical diagonalization for a 16-site cluster at zero magnetic field. We have taken the Hamiltonian $H_{\mathrm{SU(2)}}+H_{\mathrm{DM}}$ with $J_1/J_2=-3.24$, $J'/J_2=0.583$, $J_2=35.1$~K for (a) $D_{\rm u}/|J_1|=0.2$, (b) $D_{\rm u}/|J_1|=0.21$, (c) $D_{\rm u}/|J_1|=0.215$, and (d) $D_{\rm u}/|J_1|=0.22$. The horizontal orange bands point to the three low-energy spin excitation levels observed in neutron scattering experiments shown in Fig.~4(a) of the main text. 
}
\label{fig:spectrum_ed}
\end{figure}

In Fig.~\ref{fig:spectrum_ed}, we present several lowest-energy eigenvalues of $H_{\mathrm{SU(2)}}+H_{\mathrm{DM}}$ computed as a function of $D_{\rm s}$ for several choices of $D_{\rm u}$ by means of the numerical diagonalization. When $D_{\rm s}=0$, there is a U(1) symmetry in the spin space. Then, the three-fold degenerate $S=1$ excited levels are split by finite $D_{\mathrm{u}}$ into a first excited level and doubly degenerate second excited levels. The second excitation energy monotonically decreases with increasing $D_{\rm u}$. Therefore, it is clear that $D_{\mathrm{u}}$ is not the only DM interaction in the material. Next, we turn on $D_{\rm s}$. The doubly degenerate levels are then split into two. We have also checked that relaxing a condition of $\bm{D}_{\mathrm{u}}\perp\bm{D}_{\mathrm{s}}$ does not drastically change the excitation spectra.
Then, adjusting three parameters $J_2$, $D_{\rm s}/|J_1|$, and $D_{\rm u}/|J_1|$, we have successfully found a reasonable parameter set of $J_2=35.1$ K, $D_{\rm s}/|J_1|=0.39$ and $D_{\rm u}/|J_1|=0.215$, where the three levels are within the different orange bands $0.20\pm0.02$, $0.38\pm0.02$, and $0.60\pm0.04$~meV observed in inelastic low-energy neutron-scattering experiments, as shown in Fig.~\ref{fig:spectrum_ed}. 

\begin{figure}
\begin{center}
\includegraphics[width=16.0cm]{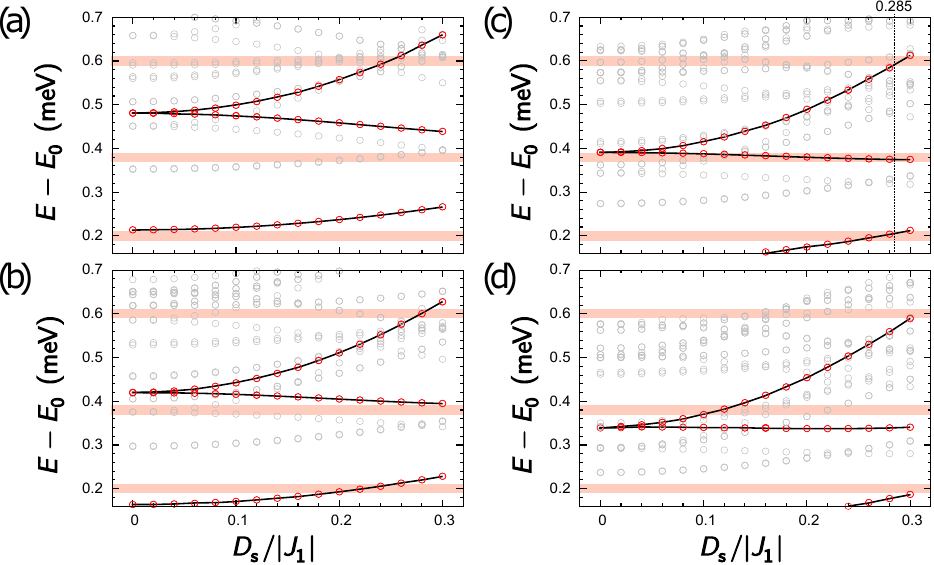}
\end{center}
\caption{Several lowest-energy levels calculated by iDMRG at zero magnetic field (red and gray circles). In particular, red circles represent twofold degenerate levels, which merge into the lowest-energy $S=1$ triplet level with an incommensurate wavenumber in the SU(2) limit.
  We have taken the Hamiltonian $H_{\mathrm{SU(2)}}+H_{\mathrm{DM}}$ with $J_1/J_2=-3.24$, $J'/J_2=0.583$, $J_2=39$~K for  (a) $D_{\rm u}/|J_1|=0.175$, (b) $D_{\rm u}/|J_1|=0.180$, (c) $D_{\rm u}/|J_1|=0.182$, and (d) $D_{\rm u}/|J_1|=0.185$. The horizontal orange bands point to the three low-energy spin excitation levels observed in neutron scattering experiments shown in Fig.~4(a) of the main text. Three black curves in each figure are guides to the eyes for trajectories of twofold degenerate levels. 
}
\label{fig:spectrum_idmrg}
\end{figure}

We have also checked finite size effects in the estimation of $J_2$, $D_{\rm s}/|J_1|$, and $D_{\rm u}/|J_1|$ by means of the iDMRG. We find a reasonable parameter set of $J_2=39$ K, $D_{\rm s}/|J_1|=0.285$ and $D_{\rm u}/|J_1|=0.182$, for which the three levels are within the different orange bands $0.20\pm0.01$, $0.38\pm0.01$, and $0.60\pm0.01$~meV, as shown in Fig.~\ref{fig:spectrum_idmrg}. Comparing $J_2$, $D_{\rm s}/|J_1|$, and $D_{\rm u}/|J_1|$ from the numerical diagonalization with those from the iDMRG, we find there is relative errors of $\sim10\%$ in $J_2$ and $D_{\rm u}/|J_1|$ and $\sim30\%$ in $D_{\rm s}/|J_1|$.

Now, we reexamine the magnetic susceptibility and the incommensurate wavevector, since we have modified $J_2$ and introduced finite $D_{\mathrm{s}}$ and $D_{\mathrm{u}}$. We first calculate a powder average of the magnetic susceptibility,
\begin{equation}
\chi(T) = \lim_{B \rightarrow 0} \frac{1}{4\pi B} \int_{0}^{\pi} \! \sin \theta d\theta \int_{0}^{2\pi} \! d\phi M(\bm{B}(B,\theta,\phi),T)
\label{eq:mag_sus}
\end{equation}
for $H_{\mathrm{SU(2)}}+H_{\mathrm{DM}}$ with the above choice of exchange parameters,
where $M(\bm{B},T)$ is a magnetization parallel to the applied magnetic field $\bm{B}$, which is expressed in spherical coordinates $(B,\theta,\phi)$. Since the angle integrations are numerically too costly, we approximate it by the average of those for $\bm{B}$ applied only along the $x$, $y$, and $z$ axes,
\begin{equation}
\chi(T) \approx \lim_{B \rightarrow 0} \frac{1}{3B} \Big[ M(\bm{B}(B,\pi/2,0),T)+M(\bm{B}(B,\pi/2,\pi/2),T)+M(\bm{B}(B,0,0),T) \Big]~.
\label{eq:mag_sus_approx}
\end{equation}
The theoretical results are shown in Fig.~S3(b) in comparison with those obtained in the SU(2)-symmetric case as well as the experimental results. We have also checked that the incommensurate wavevector is not affected by the DM interactions, as is expected since the DM interactions in the two chains of the ladder have a destructive interference.

The exchange parameters for the total Hamiltonian $H_{\mathrm{SU(2)}}+H_{\mathrm{DM}}$ are summarized as
\begin{equation}
  J_1=-114~\mathrm{K},\ 
  J_2=35.1~\mathrm{K},\ 
  J'=20.5~\mathrm{K},\ 
  D_{\mathrm{s}}=44.3~\mathrm{K},\
  D_{\mathrm{u}}=24.4~\mathrm{K}.
  \label{eq:S:parameters_total}
\end{equation}
In the main text and the rest of Supplementary Information, we adopt these values.

\section{Numerical calculations of the critical magnetic field}

\begin{figure}
\begin{center}
\includegraphics[width=8.6cm]{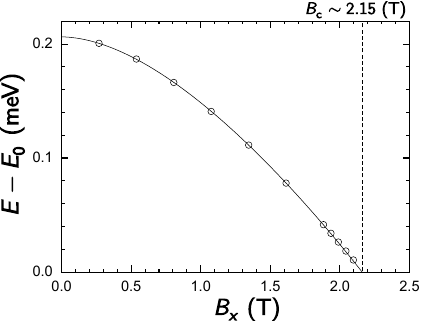}
\end{center}
\caption{
Magnetic field dependence of the first excitation energy calculated with the iDMRG method in the thermodynamic limit. The exchange parameters for $H_{\mathrm{SU(2)}}+H_{\mathrm{DM}}$ have been taken to be $J_1/J_2=-3.24$, $J'/J_2=0.583$, $J_2=39$~K, $D_{\rm u}/|J_1|=0.182$ and $D_{\rm s}/|J_1|=0.285$. The difference from the parameter set adopted in the exact-diagonalization calculations arises from finite-size effects. The dotted curve indicates the polynomial-fitted line for the lowest excitation gap. The gap collapses at a critical magnetic field $B_{\rm c} \sim 2.15$ T.
}
\label{fig:gaps_dmrg_bx}
\end{figure}

Here, we compute a critical magnetic field. We perform iDMRG calculations to reveal an evolution of several low-energy excited levels $E$ from the ground state energy $E_0$ under the magnetic field applied along the $x$, $y$, and $z$ axes. Figure~\ref{fig:gaps_dmrg_bx} shows the results for $\bm{B}$ and $\bm{D}_{\mathrm{s}}$ being parallel to $x$, in which the magnetic field lowers the first excited level most rapidly, leading to a closing of the spin gap. Then, the critical magnetic field $B_c = 2.15$~T has been obtained from an extrapolation to $E-E_0\to0$, as shown by the black curve in Fig.~\ref{fig:gaps_dmrg_bx}. This value of $B_c$ is comparable with the reported critical field 2.3~T in the experiment~\cite{Yasui14} and 2.0~T of the peak position in $d^2M/dB^2$ as shown in Fig.~3(b) of the main text.

\section{Numerical calculations of inelastic neutron-scattering spectra}

\subsection{Formulation.}

Here, we compute inelastic neutron-scattering cross-sections at $T=0$ through
\begin{equation}
I(\bm{Q},\omega) = \sum_{\alpha=x,y,z} \sum_{\beta=x,y,z} (\delta_{\alpha\beta}-Q_{\alpha}Q_{\beta}/Q^2) \left[ \sum_{\alpha'} \sum_{\beta'} R_{\alpha\alpha'} R_{\beta\beta'} \mathcal{S}^{\alpha'\beta'}(\bm{Q},\omega) \right] F(Q)^2,~
\label{eq:ins}
\end{equation}
where $R$ stands for a globally defined rotation matrix $R$, which connects the spin coordinate system with the crystallographic. 
We have also introduced the dynamical spin structure factor,
\begin{equation}
\mathcal{S}^{\alpha\beta}(\bm{Q},\omega) = -\frac{1}{\pi} {\rm Im} \bra{0} S^\alpha_{-\bm{Q}} (\omega + i \eta + E_0 - H)^{-1} S^\beta_{\bm{Q}} \ket{0}~,
\label{eq:dynamical}
\end{equation}
with the magnetic form factor $F(Q)$ of Cu and the ground-state wavefunction $\ket{0}$ and energy $E_0$. Here, the Fourier transform $S^{\alpha}_{\bm{Q}}$ of spin operators in a single two-leg ladder with the number $2L$ of spins is defined by 
\begin{equation}
S^{\alpha}_{\bm{Q}} = \frac{1}{\sqrt{2L}} \sum_{\sigma=\pm} \sum_{\ell=0}^{L-1} e^{i \bm{Q} \cdot \bm{d}_{\sigma,\ell} } S^{\alpha}_{\sigma,\ell}~,
\end{equation}
where the vectors $\bm{d}_{\sigma,\ell}$ indicates the positions of Cu atoms in the ladder and are given by 
\begin{equation}
\bm{d}_{\sigma,\ell} = \left\{ \begin{matrix} 
\frac{\ell}{2} \bm{b} & (\sigma=+,~\ell=even ) \\
\frac{\ell}{2} \bm{b} + \bm{d}_1 & (\sigma=-,~\ell=even ) \\
\frac{\ell-1}{2} \bm{b} + \bm{d}_2 & (\sigma=+,~\ell=odd ) \\
\frac{\ell-1}{2} \bm{b} + \bm{d}_1 + \bm{d}_2 & (\sigma=-,~\ell=odd )
\end{matrix} \right.
\end{equation}
with $\bm{d}_{1}=(-4.669,0.1337,1.473)$~\AA~and $\bm{d}_{2}=(0.5849,2.545,1.627)$~\AA~\cite{Solodovnikov97}. An inversion operation and a $2_1$ screw operation about the $b$ axis on the ladder generate the other three ladders in the unit cell. Therefore, as far as interladder correlations are ignored, $S^{\alpha\beta}(\bm{Q},\omega)$ per ladder both for $\bm{Q}=(0,Q_b,0)$ and after the powder averaging is independent of the ladder index.

For numerical calculations of $\mathcal{S}^{\alpha\beta}(\bm{Q},\omega)$, we perform a block continued-fraction expansion based on block Lanczos method~\cite{Cullum74} and the Schur complement, which are straightforward extensions of a continued fraction expansion based on the Lanczos algorithm~\cite{Haydock72,Gagliano87,Dagotto94}.
We adopt $\eta = 0.016J_2$ for Figs.~4(b) and 4(c) and $\eta = 0.043J_2$ for Fig.~4(e) in the main text.

\subsection{Powder averaging.}

The powder average of neutron scattering spectra is given by the angle average,
\begin{eqnarray}
  I_{\mathrm{p.a.}}(Q,\omega) & = & \frac{1}{4\pi} \int_{0}^{\pi} \! \sin \theta d\theta \int_{0}^{2\pi} \! d\phi ~ I(\bm{Q}(Q,\theta,\phi),\omega)~.
\end{eqnarray}
with the spherical coordinates $(Q,\theta,\phi)$ of $\bm{Q}$.
In the current case of the finite ladder, by taking the $b$ axis in the direction to the north pole $\theta=0$, we can write the above Eq.~as
\begin{eqnarray}
  I_{\mathrm{p.a.}}(Q,\omega)&=& \frac{1}{2k_0+1} \sum_{k=-k_0}^{k_0} \frac{1}{2\pi} \int_0^{2\pi} \! d \phi I \left( \bm{Q}\left(\sqrt{Q^2_{~}- |\bm{Q}_b^{(k)}|^2 },\pi/2,\phi \right)+\bm{Q}_b^{(k)},\omega \right),~
\end{eqnarray}
with  $\bm{Q}_b^{(k)}= (4\pi k/L)(\bm{b}/b)$, where $k_0$ is an integer determined by $| \bm{Q}_b^{(k_0)} | \leq Q < | \bm{Q}_b^{(k_0+1)} |$ in the range $0\le k_0 \le L/2-1$.

\subsection{Connecting the spin coordinate frame to the crystallographic.}

In this subsection, we fix $L=8$ unless otherwise noted. The neutron-scattering cross-sections actually depend on how the spin and crystallographic coordinate frames are related to each other, as is clear from Eq.~(\ref{eq:ins}). Now we make the following observation by a close look at Fig.~4(a): The intensity at $Q=|\bm{Q}^{(3)}_b| \sim 0.9~\mathrm{\AA}^{-1}$ is much smaller than that at $Q=|\bm{Q}^{(1)}_b|\sim0.3~\mathrm{\AA}^{-1}$, although the former can be as large as the latter since $\bm{Q}^{(3)}_b$ is related to $-\bm{Q}^{(1)}_b$ by the reciprocal lattice vector $(2\pi/b)(\bm{b}/b)$.
Therefore, we adjust the rotation matrix $R$ in Eq.~(\ref{eq:ins}) so that the ratio,
\begin{equation}
 \frac{\int_0^{1{\rm meV}} \! d\omega I(\bm{Q}^{(3)}_b,\omega)}
     {\int_0^{1{\rm meV}} \! d\omega I(\bm{Q}^{(1)}_b,\omega)}
\end{equation}
is minimized. 
The solution has been obtained by the simplex method~\cite{Numerical_recipes} as the Euler angles
\begin{equation}
  \phi=0.542\pi,~~\theta=0.724\pi,~~\psi: \mathrm{arbitrary},
\end{equation}
for b-c-b (z-x-z) type rotations.

\subsection{Interpolation of the results at discrete wavevectors.}
\begin{figure}
\begin{center}
\includegraphics[width=8.6cm]{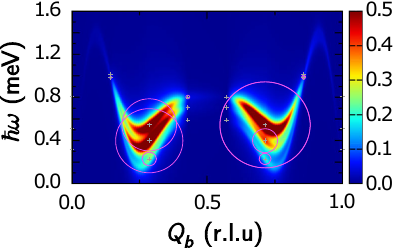}
\end{center}
\caption{
Inelastic neutron-scattering cross-section along the specific cut $\bm{Q}=(0,Q_b,0)$. The raw data are depicted by $+$ symbols at the discrete energy levels with the associated magenta circles whose radii represent the relative intensities. The results obtained by polynomial interpolations are shown in the color map for comparison and in Fig.~4(c). 
}
\label{fig:spline}
\end{figure}

The continuous spectra shown in Fig.~4(c) have been obtained by polynomial interpolations of three lowest eigenenergy spectra on a 28-site cluster with the parameter set $J_1/J_2=-3.24, J_2=39~\mathrm{K}, J'/J_2=0.583, D_{\mathrm{s}}/|J_1|=0.285 ~\mathrm{and}~ D_{\mathrm{u}}/|J_1|=0.182$ estimated by iDMRG shown in Fig.~\ref{fig:spline}.
  We assume the polynomial functions take minima at $Q_b=0.25~(0.75)$ r.l.u. to ensure the consistency with the iDMRG result on the incommensurate wavevector in this particular parameter set. We have also performed linear interpolations of the intensities at discrete $Q_b$'s.

\section{Calculations of electron spin resonance spectra}

Here, we deal with the absorbed microwave power~\cite{Slichter90} in electron spin resonance experiments.
The absorption is given by 
\begin{equation}
\mathcal{I}(\omega,\bm{B},T) \propto \mu^{-1}_{0}B^2\omega \sum_{\alpha} S^{\alpha\alpha}_{\bm{B},T}(\omega,\bm{Q}=\bm{0}),~
\end{equation}
with the permeability $\mu_0$ of the vacuum, where we have introduced the diagonal part of the same dynamical spin structure factor as in the previous section but for $\bm{Q}=\bm{0}$ under a magnetic field $\bm{B}$ at a finite temperature $T$, 
\begin{equation}
S^{\alpha\alpha}_{\bm{B},T}(\omega,\bm{0}) =  -\frac{1}{\pi} {\rm Im} \sum_{k=0}^{L/2-1} \sum_{n} e^{-\frac{E_{\bm{B},k,n}}{T}} \bra{\bm{B},k,n} S^\alpha_{\bm{0}} (\omega + i \eta + E_{\bm{B},k,n} - H )^{-1} S^\alpha_{\bm{0}} \ket{\bm{B},k,n}, 
\end{equation}
where $\ket{\bm{B},k,n}$ and $E_{\bm{B},k,n}$ are the $n$th excited state and the associated eigenenergy of the total Hamiltonian $H$ given by Eq.~(\ref{eq:H}) with the wavenumber $k$ along the chain under the magnetic field $\bm{B}$. At $T=1.6$~K,  which is comparable to the spin gap at $B=0$, transitions among a few low-energy excited states at the incommensurate wavevector and at zero wavevector largely contribute to the spectra. 
We take the same 16-site cluster and perform the numerical exact diagonalization with $\eta = 0.007J_2$. The results on the spectra are displayed in Fig.~5(c). 

\section{Experiments}
\subsection{Synthesis of polycrystalline Rb$_2$Cu$_2$Mo$_3$O$_{12}$ samples}
Polycrystalline samples of Rb$_2$Cu$_2$Mo$_3$O$_{12}$ were prepared through a standard solid-state reaction: RbCuO$_3$, CuO, and MoO$_3$ were mixed with a proper molar ratio in a grove box with inert gas, and the mixtures were pressed into pellets. The pellets were sintered at 480$^\circ$C for 72 h in air. After regrinding and pelletizing, the same heat treatment was repeated several times. No impurity phases were detected in the X-ray diffraction patterns of the finally obtained samples.

\subsection{Measurements on macroscopic physical properties}
The magnetization and magnetic susceptibility measurements were carried out with a capacitive Faraday magnetometer down to 0.06~K in a zero-field cooling process, by using a $^3$He-$^4$He dilution refrigerator. In the measurements, a field gradient of 3~T/m was applied to the samples. The specific heat was measured by the thermal relaxation method. The dielectric constant $\varepsilon$ was measured for the rectangular sample plates (typical sizes of $\sim5.0\times4.0\times0.8$~mm$^3$), to which the electrodes were attached with silver paint, by using an LCR meter (Agilent E4980A) with a frequency of 10 kHz. The electric polarization was measured from the data of the pyroelectric current using an electrometer (Keithley 6517 A).

\subsection{Set up of neutron scattering experiments}
In the low-energy neutron scattering experiments using AMATERAS~\cite{AMATERAS}, the speed of the monochromating disk chopper was fixed at 300 Hz and the other disk choppers were fixed at appropriate conditions to achieve necessary incident energies and resolutions. In the high-energy neutron scattering experiments using 4SEASONS~\cite{4SEASONS}, the speed of the Fermi chopper was fixed at 150 Hz. The data were analyzed using the software UTSUSEMI~\cite{Utsusemi}. 

\subsection{Set up of neutron diffraction experiments}
The measurements were carried out on the DMC instrument~\cite{Schefer90} at the Swiss Spallation Neutron Source (SINQ), Paul Scherrer Institute, Switzerland. 15~g of Rb$_2$Cu$_2$Mo$_3$O$_{12}$ powders was loaded into an aluminium can, and installed into the variable temperature insert of a vertical field cryomagnet. 

\subsection{Electron spin resonance experiments}
Electron spin resonance (ESR) measurements were performed by utilizing a superconducting magnet (Oxford Instruments), a vector network analyzer (MVNA, ABmm), and laboratory-built transmission-type ESR cryostat at the Center for Advanced High Magnetic Field Science in Osaka University. The measurement temperature was controlled by the variable temperature insert of the superconducting magnet.

\subsection{$g$ factor of polycrystalline Rb$_2$Cu$_2$Mo$_3$O$_{12}$}
\begin{figure}
\begin{center}
\includegraphics[width=15cm]{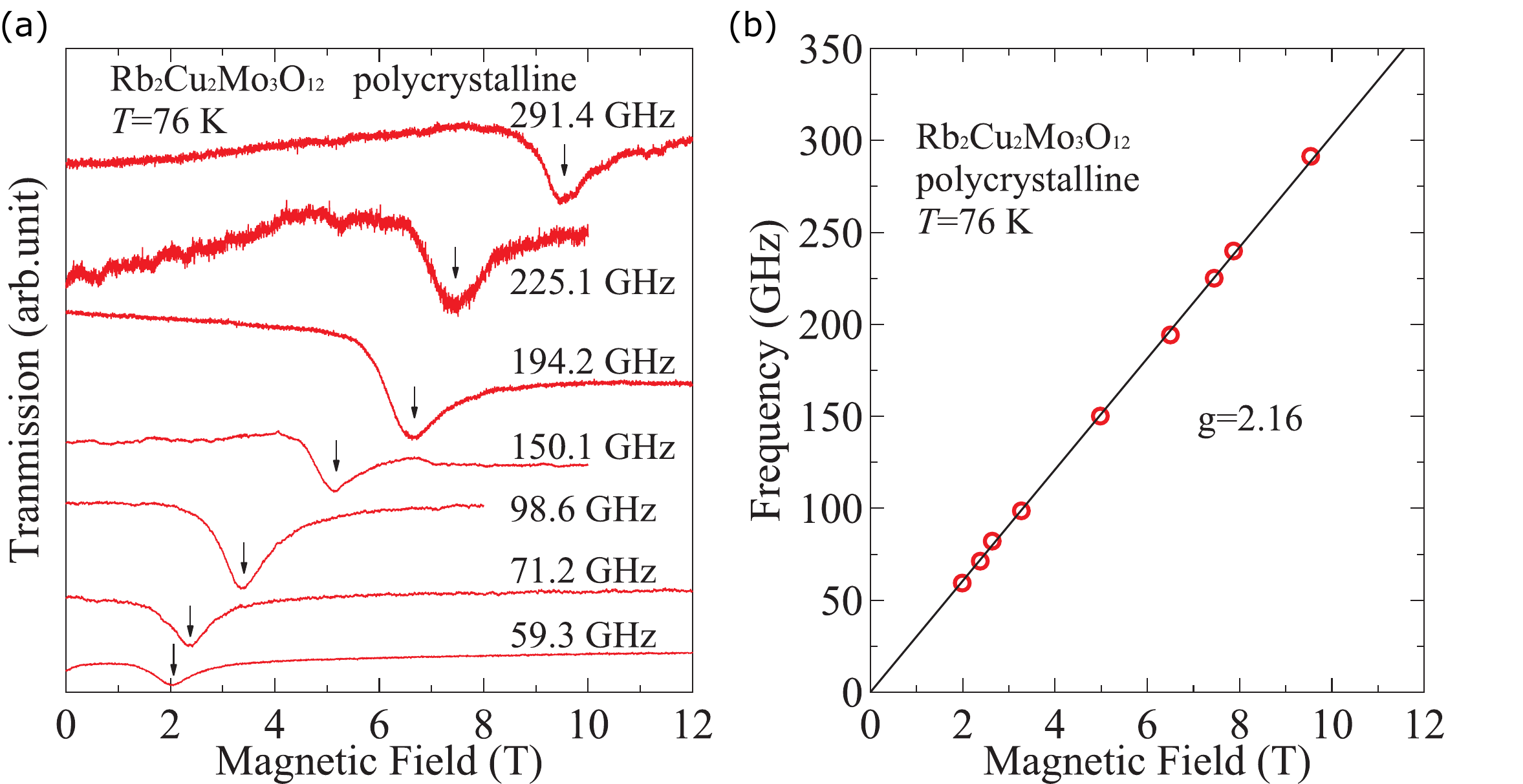}
\end{center}
\caption{
Electron spin resonance spectra of polycrystalline Rb$_2$Cu$_2$Mo$_3$O$_{12}$. 
(a) Experimental transmission spectra at 76 K for designated frequencies.
(a) Magnetic field dependence of the resonance frequency. 
}
\label{fig:esr_g}
\end{figure}

We revisit the estimation of $g$ factor of polycrystalline Rb$_2$Cu$_2$Mo$_3$O$_{12}$ by use of the electron spin resonance spectra with high magnetic field $B>2$ T. Broad but clear resonances appear in the experimental transmission spectra at 76 K (Fig.~\ref{fig:esr_g}(a)), and the resonance frequency monotonically increases with increasing magnetic field (Fig.~\ref{fig:esr_g}(b)). By use of the linear fitting of the data in the frequency range $f=[59.3:291.4]$ GHz, the $g$ factor is estimated to be $g=2.16\pm0.01$, which is larger than the reported value 2.03~\cite{Hase04}.

\end{document}